\documentclass[aps,reprint,twocolumn,superscriptaddress]{revtex4-2}

\usepackage[T1]{fontenc}
\usepackage{graphicx}
\usepackage[hidelinks,colorlinks=true,linkcolor=blue,citecolor=blue,urlcolor=blue]{hyperref}
\usepackage{amsmath}
\usepackage{bbold}
\usepackage{placeins}
\usepackage{multirow}
\usepackage[table,xcdraw]{xcolor}
\newcommand{\ket}[1]{| #1 \rangle}
\newcommand{\acomment}[1]{}
\newcommand{\citefill}[1]{\textcolor{black}{\textbf{[]}}}

\definecolor{section}{RGB}{0,0,0}

\begin{document}

\preprint{}

\title{Leakage-protected idle operation of a triangular exchange-only spin qubit}

\author{Joseph D. Broz}
\email{jdbroz@hrl.com}
\affiliation{HRL Laboratories, LLC, 3011 Malibu Canyon Road, Malibu, California 90265, USA}
\author{Jesse C. Hoke}
\affiliation{HRL Laboratories, LLC, 3011 Malibu Canyon Road, Malibu, California 90265, USA}
\author{Edwin Acuna}
\affiliation{HRL Laboratories, LLC, 3011 Malibu Canyon Road, Malibu, California 90265, USA}
\author{Jason R. Petta}
\affiliation{HRL Laboratories, LLC, 3011 Malibu Canyon Road, Malibu, California 90265, USA}
\affiliation{Department of Physics and Astronomy, University of California -- Los Angeles, Los Angeles, California 90095, USA}
\affiliation{Center for Quantum Science and Engineering, University of California -- Los Angeles, Los Angeles, California 90095, USA}

\date{\today}

\begin{abstract}
    We characterize the coherence of a triangular exchange-only (EO) spin qubit operated at a leakage-protected idle (LPI) point. The triangular geometry enables independent control of all three pairwise exchange interactions, and the LPI condition occurs when these couplings are turned on simultaneously and tuned to equal strength. In this configuration, the exchange interaction induces an energy gap $E_g=3J/2$ that suppresses leakage from the computational subspace while leaving the qubit state unaffected. We develop procedures to calibrate the LPI point and measure $E_g$, and use these to characterize the qubit dephasing time over a broad range of gap energies. While operating with large always-on exchange couplings exposes the qubit to charge noise, we find that $\tilde{T}_2^*$ still exceeds that of conventional exchange-only spin qubits for $E_g/h < 60$ MHz. The precise control of simultaneous, all-to-all connected exchange demonstrated here presents a natural path towards improving the performance of EO qubits and also enables novel qubit encodings.
\end{abstract}

\maketitle


 Spins confined in gate-defined semiconductor quantum dots (QDs) have emerged as a promising platform for quantum information processing \cite{hanson2007spins,burkard2023}. A key advantage of this technology is its scalability: QDs have small footprints, large charging energies \cite{Yang2020,Petit2020,Huang2024}, and are broadly compatible with established semiconductor fabrication techniques \cite{Ansaloni2020,ha2022,Zwerver2022,Neyens2024}. However, control of an individual spin requires engineering its local magnetic field and/or applying a microwave drive \cite{tokura2006,Pioro2008}, both posing challenges for large-scale integration. 

Alternatively, a single qubit can be encoded in the collective spin state of three exchange-coupled QDs, enabling universal control using only baseband voltage modulation of the exchange interaction \cite{DiVincenzo2000,Andrews2019}. The \textit{exchange-only} (EO) qubit encoding eliminates the need for magnetic fields and microwaves but is vulnerable to leakage, where information escapes the encoded subspace \cite{ladd2012,Kerckhoff2021}. Leakage errors are particularly costly because they cannot be corrected directly using conventional protocols and instead must be mapped onto a standard qubit error (bit or phase flip) using ancillary qubits and multi-qubit logic \cite{kempe2001-2}.

 While leakage errors can be reduced through improved epitaxy --- for example, by increasing the isotopic purity of the silicon quantum well \cite{eng2015} --- they can also be dynamically suppressed by activating multiple pairwise exchange interactions simultaneously \cite{Weinstein2005}. Recent experiments have shown that \textit{simultaneous exchange} opens up an energy gap $E_g$, that penalizes transitions out of the qubit subspace \cite{2-Jexchange}. These demonstrations, however, have thus far been limited to 2-$J$ exchange, where only two interactions are active at a time \cite{2-Jexchange,heinz2024}. With 2-$J$ exchange, leakage suppression is only available during active qubit manipulation and not during idling periods, which account for a significant fraction of processing time \cite{Andrews2019,Weinstein2023}. In contrast, by activating all three exchange interactions --- 3-$J$ exchange --- a finite energy gap can be opened while preserving the qubit state \cite{Weinstein2005}. This occurs when all three exchange interactions are tuned to equal strength, defining a leakage-protected idle (LPI) ``point'' of operation.

Here, we demonstrate leakage suppresion of an EO qubit operated at the LPI point. We outline procedures for calibrating the LPI and for determining the magnitude of the induced gap $E_g$. Using these methods, we measure the free evolution of the qubit over a wide range of $E_g$. Because exchange is controlled electrically --- by tuning the overlap of interdot wavefunctions --- stronger exchange necessarily increases the qubit's sensitivity to charge noise \cite{Reed2016}. Accordingly, we observe a strong enhancement of dephasing at large exchange couplings. However, for moderate couplings, when $E_g/h < 60$ MHz, we find that the dephasing time $\tilde{T}_2^*$ actually exceeds that of conventional EO spin qubits, which we attribute to the reduction of leakage-induced dephasing at the LPI point.


\begin{figure*}
\centering
\includegraphics[width=0.99\textwidth]{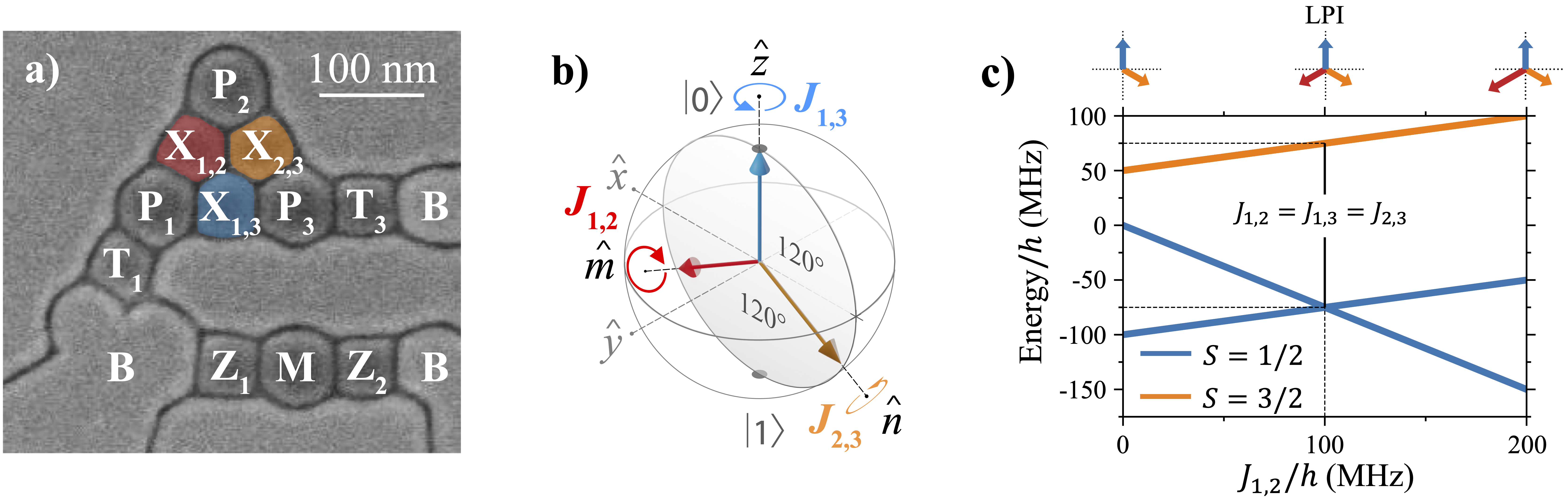}
\caption{\label{fig:device} \textbf{Simultaneous exchange coupling of three spins}.
(a) SEM image of a device similar to the one measured. (b) An isolated pairwise exchange interaction generates a qubit rotation about one of the axes $\hat{n}$, $\hat{m}$, or $\hat{z}$, whereas simultaneously activating multiple interactions generates a rotation about some vector sum of these axes. (c) Eigenergies of Eq.~\ref{eq:hamiltonian} plotted as a function of $J_{1,2}/h$ with $J_{1,3}/h$ = $J_{2,3}/h=100$ MHz and $B = 0$. The orange curve corresponds to the $S\mathbin{=}3/2$ leakage quadruplet and the blue curves correspond to the two degenerate $S\mathbin{=}1/2$ doublets used to encode the qubit. With equal exchange couplings ($J_{1,2} = J_{1,3} = J_{2,3} = J$; denoted by the black vertical line), there is no qubit rotation, but an energy gap $E_g = 3J/2$ is induced between the $S\mathbin{=}1/2$ and $S\mathbin{=}3/2$ subspaces.  The energy gap suppresses leakage from local magnetic field fluctuations, yielding an LPI point.}
\end{figure*}

Measurements are performed on a triangular triple QD device consisting of a single layer of etch-defined gate electrodes fabricated on an isotopically enriched $^{28}$Si/SiGe heterostructure (800 ppm $^{29}$Si), as shown in Fig.~\hyperref[fig:device]{1(a)} \cite{ha2022,acuna2024}. Plunger gates ($\textrm{P}_i$) primarily control the chemical potentials of the dots formed beneath them, and are each biased to confine a single electron. Exchange gates ($\textrm{X}_{i,j}$) primarily control the interdot tunnel couplings. A dynamic exchange interaction between dots $i$ and $j$ is implemented by pulsing the virtual exchange gate voltage $\tilde{V}_{X_{i,j}}$, composed of a linear combination of physical gate voltages chosen to cancel cross-coupling between gates to first order \cite{2-Jexchange,Hensgens2017fermi-hubbard,mills2019,hsiao2020}.

The EO qubit is encoded in the collective three-electron spin state. Under $S$, the total electron spin, the associated Hilbert space decomposes into two $S\mathbin{=}1/2$ Zeeman doublets and one $S\mathbin{=}3/2$ quadruplet. The doublets are used to define the qubit, with the encoding $\ket{0}\mathbin{=}\ket{S_{13}\mathbin{=}0,S\mathbin{=}1/2,m_S\mathbin{=}\pm 1/2}$ and $\ket{1}\mathbin{=}\ket{S_{13}\mathbin{=}1,S\mathbin{=}1/2,m_S\mathbin{=}\pm 1/2}$, where $S_{13}$ is the total spin of the electrons in dots 1 and 3, and $m_S$ is left as a ``gauge'' degree of freedom \cite{burkard2023}. Qubit initialization and readout both rely on Pauli spin blockade between dots 1 and 3, which only distinguishes $S_{13}$ \cite{Andrews2019,blumoff2022}. The particular choice of gauge ($m_S=\pm1/2$) is randomly assigned during initialization and left unresolved by measurement \cite{blumoff2022}. Furthermore, readout determines only the occupation of $\ket{0}$ and does not discriminate between $\ket{1}$ and the spin-symmetric $S\mathbin{=}3/2$ leakage states (which all have $S_{13}=1$) \cite{burkard2023}.

The exchange interaction is described by the Hamiltonian:
\begin{equation} \label{eq:hamiltonian}
    \hat{H} = \sum\nolimits_{i<j} J_{i,j}\hat{\mathbf{S}}_i\cdot\hat{\mathbf{S}}_j,
\end{equation}
\noindent where $\hat{\mathbf{S}}_i$ denotes the spin operator for dot $i$. We refer to this as $k$-$J$ exchange, where $k$ is the number of nonzero terms in the sum. Because $\hat{H}$ is globally spin-conserving, it preserves the qubit gauge and the qubit subspace is closed under its action. In contrast, non-spin-conserving local magnetic field fluctuations arising primarily from hyperfine coupling to nearby spinful nuclei can induce transitions between $S\mathbin{=}1/2$ and $S\mathbin{=}3/2$ states, causing leakage errors \cite{ladd2012,Kerckhoff2021}.

Within the (gauge-dependent) qubit subspace, each individual term in $\hat{H}$ generates a rotation about an axis $\hat{z}$, $\hat{n}$, or $\hat{m}$ distributed uniformly over the $xz$-plane of the Bloch sphere [Fig.~\hyperref[fig:device]{1(b)}]. Simultaneous exchange occurs when $k>1$, and generates a rotation about an axis aligned along $\vec{r}= J_{1,2}\hat{m} + J_{2,3}\hat{n} + J_{1,3}\hat{z}$. In particular, $\vec{r}=0$ when $J_{1,2} = J_{1,3} = J_{2,3} = J$ and no rotation occurs. If $J\ne 0$, we refer to this condition as the LPI point, under which the Hamiltonian reduces to $\hat{H}=J\hat{S}^2/2 - 9/8$, inducing a gap of $E_g = 3J/2$ between the $S\mathbin{=}1/2$ and $S\mathbin{=}3/2$ subspaces [Fig.~\hyperref[fig:device]{1(c)}]. \textcolor{black}{This gap energetically protects against leakage due to off-resonant local magnetic field fluctuations by suppressing population transfer \cite{Weinstein2005}. Since hyperfine coupling strengths in these devices are typically observed to follow an inverse power law with most of the energy concentrated below several tens of kilohertz \cite{Kerckhoff2021}, even a moderately sized gap $E_g$ can significantly reduce these errors.}


\begin{figure}[h]
\centering
\includegraphics[width=0.45\textwidth]{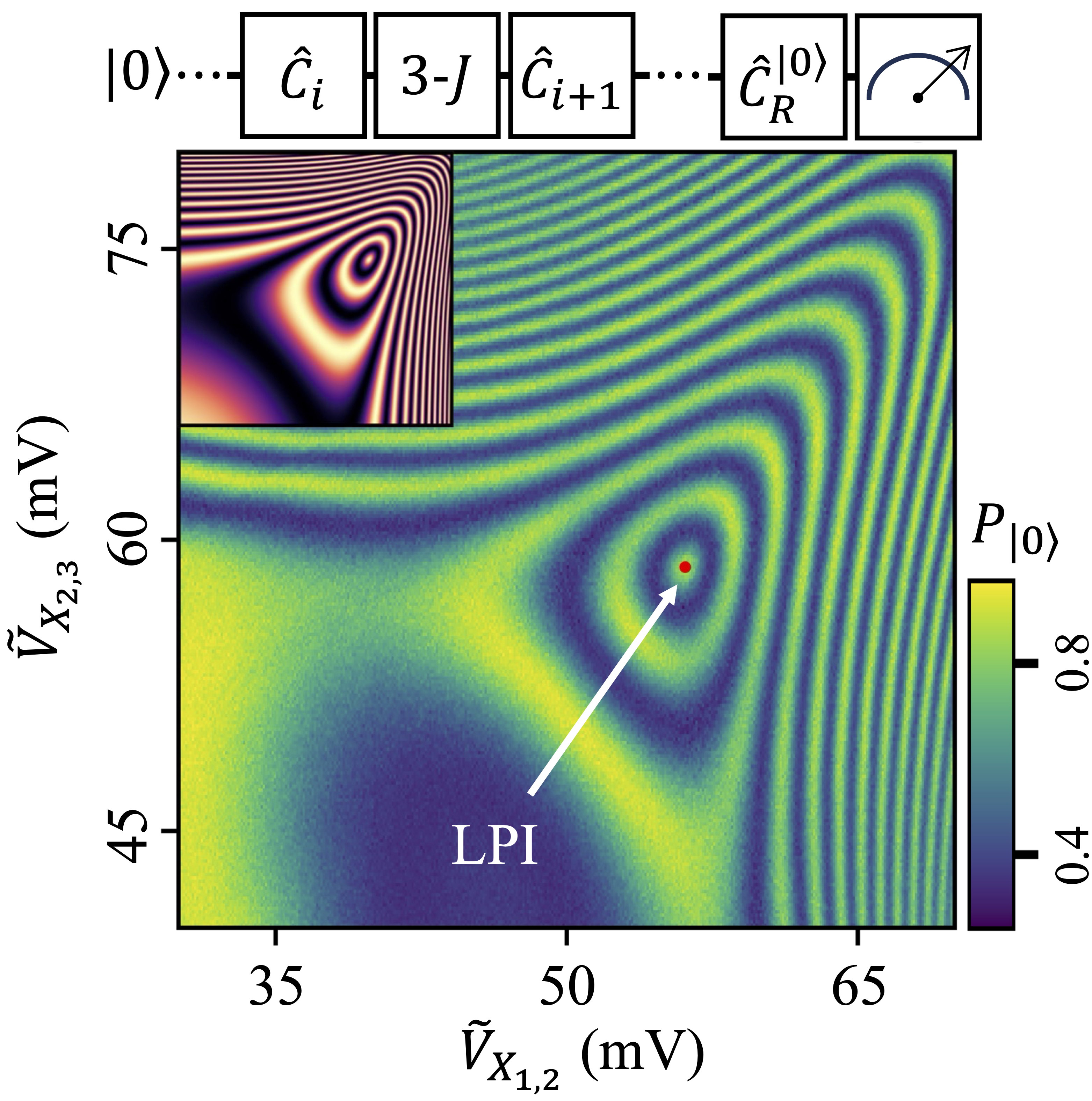}
\caption{\label{fig:exchange sweeps} \textbf{Locating the LPI point.} A 20 ns 3-$J$ pulse is interleaved within a 1-$J$ RB sequence (top diagram). One of the 3-$J$ virtual exchange gate voltages is held fixed, while the other two are swept. The red circle marks the location of the LPI point where $J=J_{1,2}=J_{1,3}=J_{2,3}\approx h\times 200$ MHz. \textcolor{black}{Inset: Simulation assuming an exponential dependence of exchange energies on gate voltages. Data are distorted relative to simulation due to a complex cross-coupled dependence of exchange energies on gate voltages,} but the LPI feature (a closed disc of high $P_{\ket{0}}$) is robust to this distortion. 
}
\end{figure}

\textcolor{black}{To locate the LPI point, we interleave a simultaneous 3-$J$ exchange pulse within a conventional 1-$J$ (only a single exchange interaction active at a time) randomized benchmarking (RB) sequence that is designed to return the qubit to $\ket{0}$ when initialized in $\ket{0}$ \cite{Andrews2019}. The measured return probability $P_{\ket{0}}$ is maximized when the 3-$J$ pulse is tuned to the LPI point—where all three exchange couplings are equal—since the pulse then implements the identity and induces no qubit rotation. We repeat this procedure many times, holding one of the virtual exchange gate voltages defining the 3-$J$ pulse fixed while sweeping the remaining two against each other.} An example sweep is shown in Fig.~\ref{fig:exchange sweeps}, with the inset displaying a corresponding simulation assuming an exponential dependence of exchange energy on gate voltages. \textcolor{black} {The data are distorted relative to the idealized simulation due to a complex cross-coupled dependence of exchange energies on gate voltages} \cite{Reed2016,coherent_multispin_exchange}. Nevertheless, the LPI point (red circle) manifests as a distinct feature --- the center of a closed disc of high $P_{\ket{0}}$ \textcolor{black}{where no rotation occurs}. The bright fringes surrounding the LPI point correspond to rotations of angle $0$ mod $2\pi$. The size of the central disc scales inversely with the 3-$J$ pulse duration, allowing us to refine the LPI estimate by increasing the length of the 3-$J$ pulse until limited by decoherence (due to hyperfine noise at low $E_g$ or charge noise at high $E_g$).

\begin{figure}
    \centering
    \includegraphics[width=0.49\textwidth]{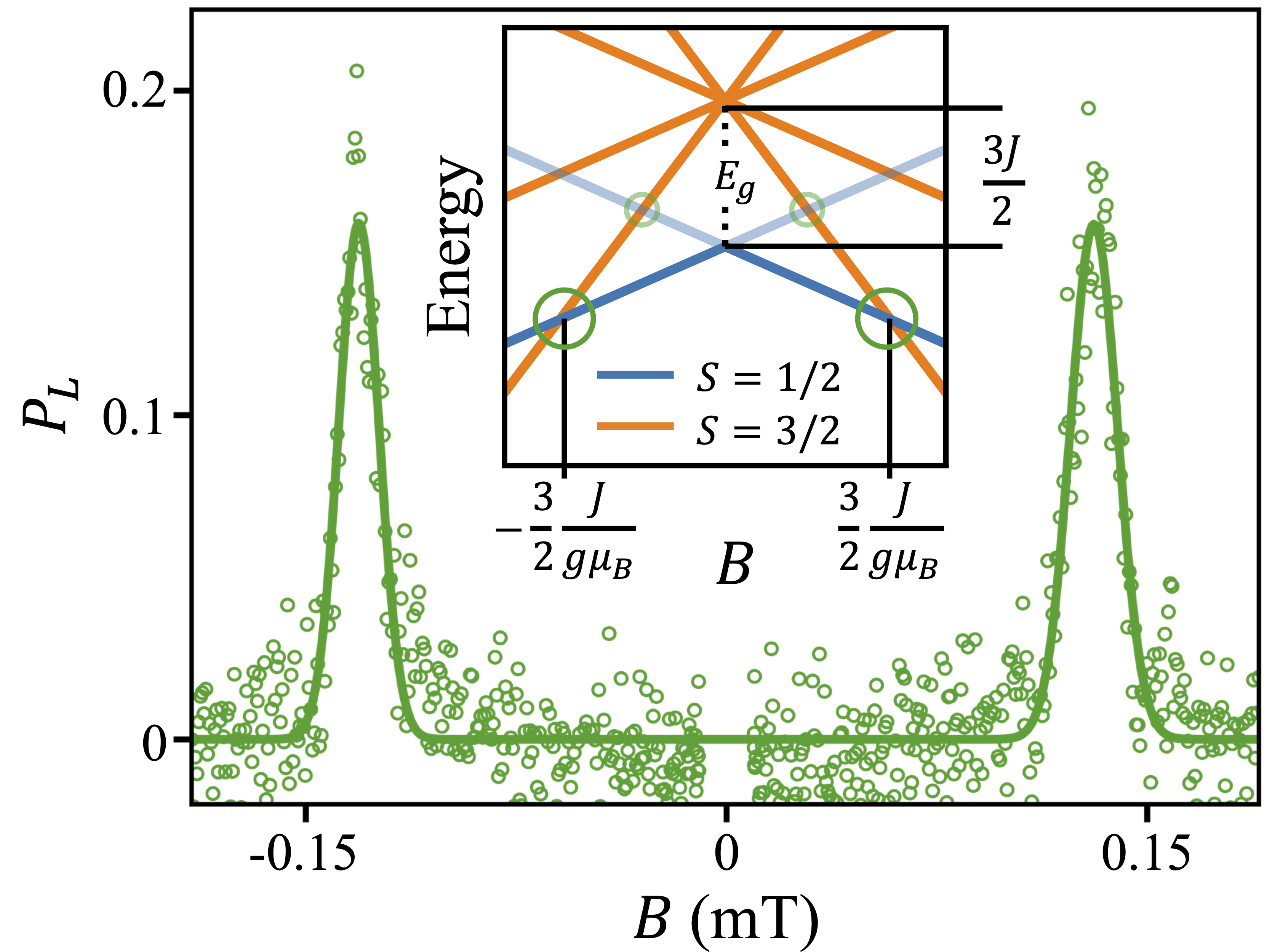}
    \caption{\label{fig:leakage spectroscopy} \textbf{EO qubit leakage spectroscopy.} Inset: EO qubit eigenenergies plotted as a function of $B$ with $J_{1,2}=J_{1,3}=J_{2,3}=J$. The green circles identify points where the $S\mathbin{=}3/2$ leakage subspace intersects with the $S\mathbin{=}1/2$ qubit subspace. \textcolor{black}{When $B$ is positive (negative), the qubit is preferentially initialized in the $m_S=+1/2$ ($m_S=-1/2$) gauge indicated by the darker shaded blue lines. Thus, only two level crossings are observed in practice (larger pair of green circles).} Main plot: Leakage probability $P_L$ measured as a function of $B$. The two peaks correspond to the level crossings circled in the inset.  The solid line is a fit to a double Gaussian from which we estimate $E_g/h=3.6\pm0.2$ MHz. The gap in data near $B=0$ is due to the ambient magnetic field (see Supplementary Information).
    }
    \end{figure}

After locating the LPI point, we measure the size of the induced gap $E_g$ using leakage spectroscopy. The relation $E_g=3J/2$ holds only in the absence of a magnetic field, where $E_g$ is single-valued due to the separate degeneracies of the $S\mathbin{=}1/2$ and $S\mathbin{=}3/2$ subspaces. Applying an external field $\vec{B}=B \hat{z}$, lifts this degeneracy, producing pairs of level crossings where the gap vanishes at $B=\pm 3J/4g\mu_B$ and $B=\pm 3J/2g\mu_B$, \textcolor{black}{corresponding to the two different qubit gauges} (Fig.~\ref{fig:leakage spectroscopy} inset) \cite{petta05}. By sweeping the magnetic field and measuring the leakage population $P_L$ at each point after a long 3-$J$ LPI pulse, we can identify these crossings as peaks in $P_L$, providing a means to estimate $E_g$. 

To actually measure $P_L$, we perform two experiments for each value of $B$: one with and one without a final $\pi$-rotation about the $x$-axis that maps $\ket{0}\rightarrow\ket{1}$. These allow us to independently measure the probabilities $P_{\ket{0}}$ (\textcolor{black}{$P_{\ket{0}\rightarrow\ket{1}}$}) that the qubit, initialized in $\ket{0}$, is found to occupy $\ket{0}$ ($\ket{1}$) after the 3-$J$ pulse. From these we compute $P_L=1-P_{\ket{0}} - \textcolor{black}{P_{\ket{0}\rightarrow\ket{1}}}$ \cite{Andrews2019}. Distinguishing leaked states from the encoded $\ket{1}$ state is essential, since decoherence during the long 3-$J$ pulse --- arising from dispersive hyperfine coupling or charge noise --- can mix $\ket{0}$ and $\ket{1}$ even without leakage. An example sweep is shown in Fig.~\ref{fig:leakage spectroscopy}, where only two peaks are visible. At finite magnetic field the qubit is preferrentially initialized in the lower-energy gauge, so the peaks at $\pm B=3J/4g\mu_B$ are not observed in practice. 

\begin{figure*}
\centering
\includegraphics[width=0.99\textwidth]{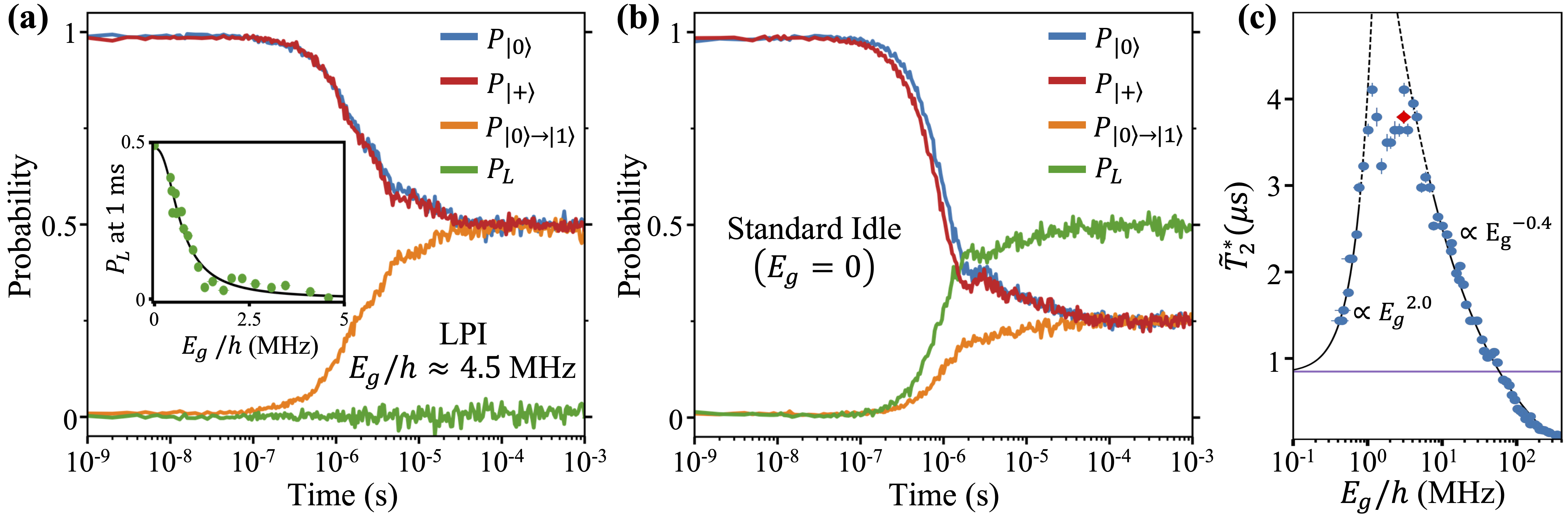}
\caption{\label{fig:lpi} \textbf{Free evolution at the LPI point.} (a) \textcolor{black}{Free evolution} at the LPI point with $E_g\approx h\times4.5$ MHz. $P_{\ket{0}}$ and $P_{\ket{+}}$ measure the probability that the qubit, initialized in the subscripted state, remains unperturbed as a function of time. $P_{\ket{1}}$ and $P_L$ measure the probabilities that the qubit, initialized in $\ket{0}$, is found to subsequently occupy either $\ket{1}$ or a leaked state, respectively. At long times, $P_{\ket{0},\ket{1}}\rightarrow 0.5$ --- consistent with a mixture over the two-dimensional qubit subspace. Inset: $P_L$ measured at 1 ms as a function of $E_g$, \textcolor{black}{with the black line a power law fit that scales approximately as $1/E_g^2$.} (b) A similar set of experiments, but now performed at the conventional (exchange-off) idle point. Here, $P_L\rightarrow0.5$, $P_{\ket{0},\ket{1}}\rightarrow 0.25$ consistent with a fully mixed state over the full eight-dimensional Hilbert space. (c) Dephasing time ($\tilde{T}_2^*$) measured at the LPI point as a function of $E_g$. The red diamond corresponds to the data in (a) and the purple line marks the $\tilde{T}_2^*$ derived from (b). Initially, $\tilde{T}_2^*$ increases as a function of $E_g$ due to suppression of leakage, but then (at $E_g/h\approx 4.5$ MHz) begins to decrease due to the mounting effects of charge noise. The black lines show power law fits (with the exponent a fitted parameter) to the low- and high-$E_g$ behavior. All data was taken at $B=0$.
}
\end{figure*}

After calibrating the LPI point and measuring the induced gap, we next characterize its performance by measuring the qubit's dynamics under free evolution. Experiments are performed with two different qubit initializations, $\ket{0}$ and $\ket{+}=(\ket{0}+\ket{1})/\sqrt{2}$. \textcolor{black}{The state $\ket{+}$ is prepared by initializing the qubit in $\ket{0}$ followed by a 1-$J$ Hadamard pulse.} For each, we measure the respective probabilities $P_{\ket{0}}$ and $P_{\ket{+}}$ that the qubit remains in its initialized state as a function of time. An example for $E_g\approx h\times4.5$ MHz is shown in Fig.~\hyperref[fig:lpi]{4(a)}. The absence of oscillations in both $P_{\ket{0}}$ and $P_{\ket{+}}$ confirms that the 3-$J$ exchange pulse does not drive rotations. 

We also perform a third \textcolor{black}{free evolution experiment} measuring \textcolor{black}{$P_{\ket{0}\rightarrow\ket{1}}$ in order to determine $P_L$}. For $E_g\approx h\times4.5$ MHz, no appreciable leakage population is observed up to at least 1 ms. For comparison, Fig.~\hyperref[fig:lpi]{4(b)} shows the corresponding measurements at the conventional (exchange-off) idle point. In this case, we observe a finite leakage population after several hundred nanoseconds, which eventually saturates to a value of 0.5 at about 100 microseconds --- consistent with a fully mixed state over the eight-dimensional Hilbert space. More generally, we expect the leakage rate to scale with the hyperfine noise power at the gap frequency $E_g/h$, which generally follows an inverse power law in these devices \cite{Kerckhoff2021}. \textcolor{black}{This trend is supported by the observed rapid suppression of leakage population with gap energy illustrated in the inset of Fig.~\hyperref[fig:lpi]{4(a)}.}

A notable feature of Figs.~\hyperref[fig:lpi]{4(a-b)} is that operation at the LPI point not only suppresses leakage, but also enhances coherence. \textcolor{black}{In Fig.~\hyperref[fig:lpi]{4(c)}, we measure the dephasing time $\tilde{T}_2^*$ as a function of $E_g$. Here,} $\tilde{T}_2^*$ is defined as the time for the average signal $\bar{P} = \langle P_{\ket{0}}(t)+P_{\ket{+}}(t) \rangle_t$ to decay to $1/e$ of its initial value, as determined numerically. This unconventional definition is chosen to capture the dephasing effects of charge noise along both the $x$- and $z$-axes of the Bloch sphere using a single parameter. At $E_g/h\approx 4.5$ MHz, we obtain $\tilde{T}_2^*\approx 3.79\pm 0.17$ $\mu$s \textcolor{black}{[red diamond in Fig.~\hyperref[fig:lpi]{4(c)}]}, compared to $\tilde{T}_2^*\approx0.85\pm0.02$ $\mu$s at $E_g=0$ (purple line). For small $E_g$, $\tilde{T}_2^*\propto E_g^2$, which we attribute to a suppression of the dephasing channel associated with leakage. At large $E_g$, where charge noise dominates, $\tilde{T}_2^*\propto 1/E_g^{0.4}$. \textcolor{black}{In both cases, these exponents are obtained from a best fit to the data.} For $E_g/h$ between 1 and 15 MHz, $\tilde{T}_2^*$ exhibits a plateau, which we attribute to dispersive hyperfine coupling. For all values of $E_g/h$ < 60 MHz, $\tilde{T}_2^*$ exceeds that measured for exchange-off idling, \textcolor{black}{as noted by the purple line in Fig.~\hyperref[fig:lpi]{4(c)}}.


 In conclusion, we have demonstrated control of simultaneous 3-$J$ exchange tuned to a leakage-protectected idle point, and established a procedure for measuring the induced energy gap $E_g$. LPI behavior was verified at $J/h\approx 3$ MHz by observing the qubit's free evolution as a function of time. The lack of oscillations in these measurements confirms the implementation of an identity operation, while the observed suppression of leakage confirms the leakage-protected property of 3-$J$ exchange. Measurements of $\tilde{T}_2^*$ over a broad range of $E_g$ further reveal that, despite increased sensitivity to charge noise, coherence times at the LPI point can still outperform conventional (exchange-off) qubit idling for $E_g/h < 60$ MHz where charge noise is not the dominant dephasing mechanism.

The EO qubit encoding inherently forms a decoherence-free subsystem (DFS) that is immune to \textit{global} magnetic field fluctuations \cite{kempe2001,DiVincenzo2000}. Operation at the LPI point complements this property by transforming the DFS into an interaction free subspace (IFS) \cite{Zhou2002} that is additionally insensitive to \textit{local} magnetic field fluctuations. A natural next step is to demonstrate transient pulsing out of the IFS to perform qubit operations, which could be realized using simultaneous 2-$J$ exchange \cite{2-Jexchange} that maintains a leakage-suppressing energy gap. More generally, this work represents the first demonstration of precise control and characterization of a simultaneous all-to-all exchange interaction in a QD device with a closed ring-like topology. In such a configuration, simultaneous exchange can give rise to a virtual tunnel current with an associated direction \cite{scarola2004} --- clockwise or counter-clockwise --- representing a new chiral quantum degree of freedom that can be exploited for novel qubit encodings \cite{hsieh2010,Srinivasa2007} or in analog quantum simulations \cite{AVISHAI2005334,ingersent2005}.

\section{Acknowledgments}
We thank John B. Carpenter for assisting with the preparation of the figures, Quantum Machines for access to the hardware used to perform the experiments (QDAC-II and OPX+), and Minh Nguyen for logistical support. The sample used in this experiment was made by the HRL device fabrication team. Research was supported by Army Research Office (ARO) grants
W911NF-24-1-0020 and W911NF-22-C-0002. The views and conclusions contained in this document are those of the authors and should not be interpreted as representing the official policies, either expressed or implied, of the ARO or the U.S. Government. The U.S. Government is authorized to reproduce and distribute reprints for Government purposes notwithstanding any copyright notation herein.


%

\clearpage
\onecolumngrid
\begin{center}

{\Large \textbf{Supplementary Information for: \\ Leakage-protected idle operation of a triangular exchange-only qubit spin qubit}}
\end{center}

\setcounter{figure}{0}
\section{1. Magnetic Field Calibration}

\begin{figure*}
	\renewcommand{\thefigure}{S\arabic{figure}}
\centering
\includegraphics[width=1\textwidth]{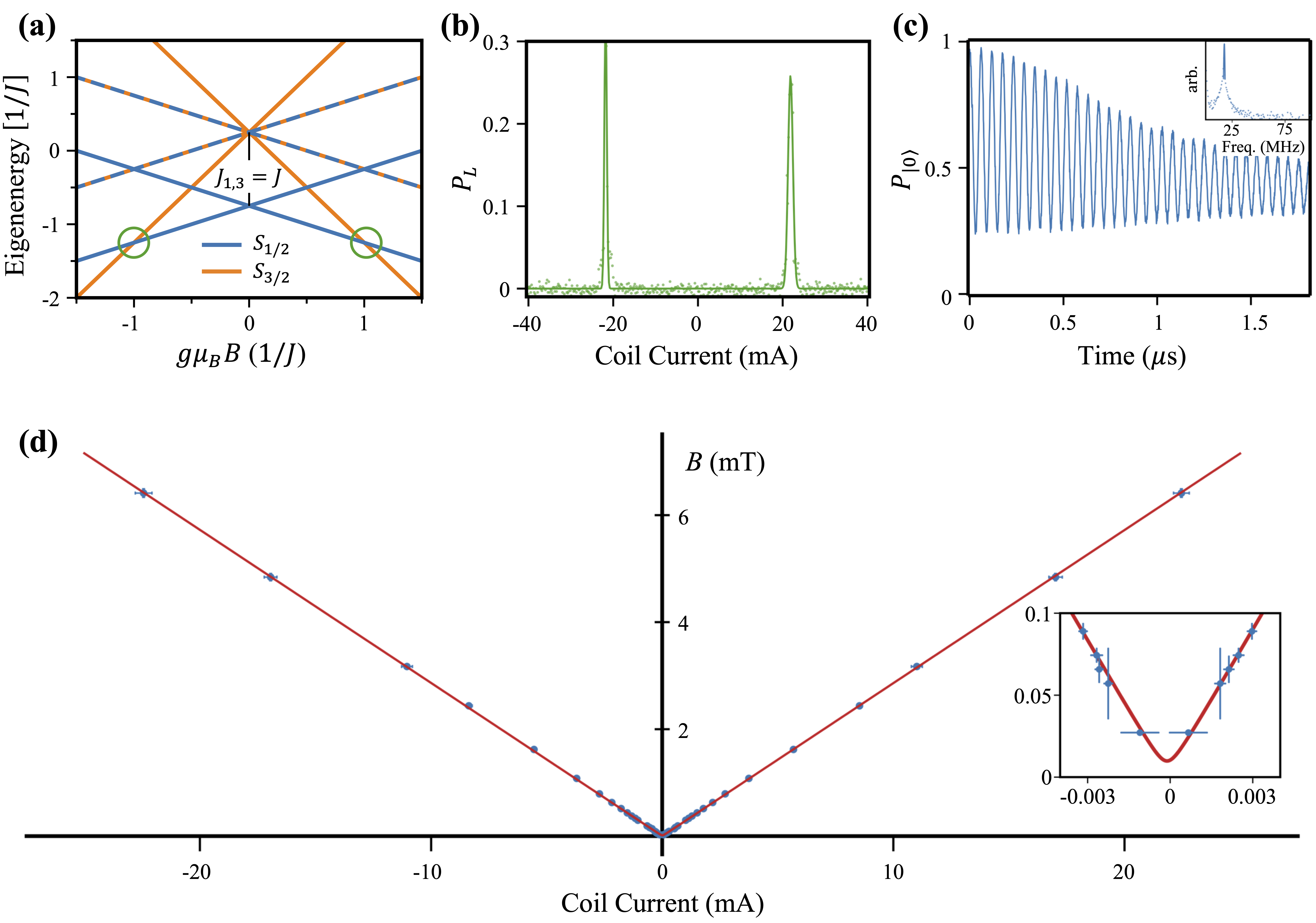}
\caption{\textbf{Magnetic field calibration.} (a) Eigenenergies of the three-electron system as a function of applied magnetic field strength with a fixed 1-$J$ exchange energy $J_{1,3}=J$ ($J_{1,2}=J_{2,3}=0$). (b) Measured leakage population as a function of magnet coil current while applying a fixed 1-$J$ exchange energy $J_{1,3}=J$. The two peaks correspond to the green circles in (a) where the $\ket{1, 3/2, \pm 3/2}$ leakage states intersect with the $S_{1/2}$ computational subspace. (c) Measured exchange oscillations used to extract the value of $J$ (in this case, $\approx$ 23 MHz as shown in the inset). (d) Using the value of $J$ measured in (c) and the location of the peaks measured in (b), we can calculate the magnetic field at the position of the dots as a function of the applied coil current (blue circles). The red curve is a fit to this data that takes into account a potential ambient magnetic field component perpendicular to the direction of the applied magnetic field.
}
\label{fig:s1}
\end{figure*}

In order to accurately to determine the magnitude of the simultaneous 3-$J$ exchange energy $J$ via leakage spectroscopy, the magnetic field $B$ seen by the quantum dots needs to be well calibrated. To do so, we first perform leakage spectroscopy measurements in the case where only one exchange energy is activated (1-$J$). As shown in the energy diagram of Fig.~\ref{fig:s1}(a), the $S_{1/2}$ and $S_{3/2}$ levels cross when $J = g \mu_B B$. This simple relationship will allow us to accurately determine the applied magnetic field $B$ by measuring $J$ via exchange oscillations.

With fixed $J_{1,3} = J$ (which is yet undetermined and is set by the plunger gate voltage $\tilde{V}_{X_{1,3}}$) and $J_{1,2}=J_{2,3}=0$, we first sweep the coil current of the superconducting magnet, which directly controls the (also yet uncalibrated) applied $B$. As seen in Fig.~\ref{fig:s1}(b), we observe two peaks in the leakage population for both positive and negative current values, which precisely correspond to the condition $J = g \mu_B B$. To determine $J$, we measure exchange oscillations between the $\ket{0}$ and $\ket{1}$ states as a function of time (Fig.~\ref{fig:s1}(c)). The Fourier transform of $P_{\ket{0}}(t)$ (inset of Fig.~\ref{fig:s1}(c)) yields a peak in frequency space that corresponds to the exchange energy $J$. From this value of $J$ we accurately calibrate the $B$ seen by the quantum dots at the currents in which the leakage is peaked. Repeating this procedure for different values of $J$ enables us to determine $B$ as a function of the magnet current (Fig.~\ref{fig:s1}(d)).

\end{document}